# Atomically flat carbon monolayer as an extremely unstable quasi-2D mesoscopic quantum system

Marina V Krasinkova*

Ioffe Institute, Politekhnicheskaya 26, St.Petersburg 194021, Russia

E-mail: marinak@mail.ioffe.ru

**Abstract.** The carbon monolayer band structure calculated in the approximation of weakly interacting π electrons corresponds to massless electron excitations known as Dirac fermions not previously observed in any other material. However, if strong Coulomb and exchange interactions between π electrons are taken into account, another picture of the π electron state emerges. These interactions result in π electron localization and electron crystal formation. The atomically flat layer can be regarded as a simplest quasi-two-dimensional mesoscopic quantum system consisting of a carbon ion plane and two π electron crystals on opposite sides of the plane. Such a system must have dielectric and pronounced diamagnetic properties and a high sensitivity to external factors distorting its electron crystals. The instability manifests itself in a tendency of the monolayer to be transformed into a more stable carbon modification with a rolled-up or wrapped-up carbon skeleton which is observed as a monolayer corrugation (or ripples). The corrugated monolayer is characterized by the presence of excited π electrons which are responsible for its physical and chemical properties. The approach can prove useful for investigation of carbon nanotubes, fullerenes, graphite, and also topological insulators and complicated quantum systems with a layered structure. Calculations of the quasi-two-dimensional quantum system faces a many-body problem and are beyond the band-structure description, which forces us to confine ourselves to purely qualitative analysis.

Keywords: graphene, 2D quantum system, electron crystal, strongly correlated electron state, unstable carbon modification.

# 1. Introduction

Interest to the carbon monolayer arose in the 1930-40s when the band structure of graphite was calculated in the framework of the tight-binding model [1] in which the hexagonal crystal lattice of the carbon monolayer proved to be responsible for an exceptional topology of the band structure radically differing from the parabolic band. Such a topology corresponds to a bad metal with massless linearly dispersing electronic excitations known as Dirac particles. During the ensuing years the band structure calculations by different methods, but in the approximation of weakly interacting electrons, confirmed the main conclusion on the dependence of the $\pi$ electron state on the crystal lattice symmetry [2-4]. However, no experimental confirmation of these calculations was obtained (i) because of unsuccessful attempts to synthesize a graphite monolayer, and (ii) because the estimations of the 2D crystal stability showed that a long-range order could not exist in a purely 2D system due to its sensitivity to density fluctuations [5, 6].

Attention to the carbon monolayer increased when such carbon modifications as fullerenes and nanotubes based on a wrapped up and rolled up carbon monolayer were discovered. Later the flat carbon monolayer was obtained from graphite [7,8] and first experimental data on its properties were reported [9].These publications were regarded as the first experimental confirmation of the existence of a 2D carbon monolayer with the carriers similar to massless Dirac particles, not previously observed in any other material.

In spite of a vast body of experimental and theoretical data on properties of the carbon monolayer available at present, no answers to many important questions have been found so far. First, it is still unclear why quasiparticles in graphite that consists of carbon monolayers with the interlayer distance (3.37Å) exceeding in more than 2 times the distance between carbon atoms in the monolayer (1.42Å) have a non-zero effective mass, while the low-energy quasiparticle mass in the carbon monolayer is zero. In other words, how a weak interlayer interaction in graphite (as compared with a strong intralayer one) is able to change the state of carriers so much that massless Dirac particles become ordinary fermions. Second, it is not clear why, as observations of the Shubnikov-de Haas effect and cyclotron resonance study of the carbon monolayer show [10, 11], the carrier cyclotron mass differs from zero. The cyclotron mass can, of course, differ from the effective mass of the particles found by other methods, but this difference relates only to the mass magnitude rather than to the nature of particles. Third, as follows from experimental data [12], a point distortion of the monolayer by, for instance, adsorption of one hydrogen atom on a graphite surface, causes an electron density redistribution in definite crystallographic directions at the length of about 20-25 lattice constants. This indicates that some long-range interactions, hitherto unknown and not taken into account, are present in the surface layer.

This list of hardly explainable experimental facts can be enhanced by the anomalous quantum Hall effect observed at room temperature [13], ballistic conductivity [14], universal conductance fluctuations [15], and out-of-plane deformations in both substrate-supported and suspended carbon monolayers [16]. So the question arises as to whether the band structure calculated in the approximation of weakly interacting particles [1-4] adequately reflects the real state of $\pi$ electrons in the atomically flat carbon monolayer and whether the obtained experimental data relate to a really flat carbon monolayer.

This paper suggests a new approach to the atomically flat carbon monolayer resulting from the consideration of Coulomb and exchange interactions between $\pi$ electrons. It follows from this approach that (i) the atomically flat 2D carbon modification is extremely sensitive to the factors that distort its $\pi$ electron subsystem, and (ii) if such a flat carbon monolayer were stable, it would have properties of a dielectric and a diamagnetic rather than of a semiconductor or a bad metal. It also follows from the approach that the calculations of properties of such a carbon modification faces a many-body problem and are beyond the band-structure description. For this reason we are forced to confine ourselves to a purely qualitative analysis.

## 2. Results and Discussion

### 2.1. *Formation of $\pi$ electron crystals in the atomically flat carbon monolayer.*

All $\pi$ electrons in the atomically flat carbon monolayer participate in the formation of identical $\pi$ bonds, which increases the bond multiplicity between atoms and reduces the interatomic distance. The $\pi$ bonds are similar to the bonds in carbon compounds with the resonance of structures. A distinguishing feature of $\pi$ bonding in the carbon monolayer is the formation of three $\pi$ bonds by each $\pi$ electron. This means that each $\pi$ electron is shared by the atom to which it belongs and three neighboring carbon atoms. In this case each $\pi$ electron is so tightly bound with three neighboring carbon atoms that it is localized in the space surrounded by these three atoms. In other words, it is localized at its own atom in the $2p_z$ state which overlaps with the 2p$_z$ states of three neighboring atoms. Since all π electrons localized in this manner prove to be at a short distance from each other (~1.42Å) and are out of the carbon skeleton plane, a strong unscreened Coulomb interaction arises between them (~10 eV between each two neighboring electrons).

The important role of the Coulomb interaction between π electrons in properties of aromatic carbon compounds was discussed for the first time in the 1970s when attempts to explain rich optical spectra of benzene and naphthalene were made [17, 18]. The hypothesis was put forward that the Coulomb interaction between neighboring π electrons in benzene had to lead to their correlated state in which they were arranged in quasi-1D Wigner crystals.

Each π electron forming three π bonds in the flat carbon monolayer experiences the Coulomb repulsion from three neighboring π electrons. Because of the repulsion these π electrons are forced to be at a maximum possible distance from each other and be arranged alternately on both sides of the carbon skeleton. In other words, the Coulomb interaction between neighboring π electrons leads to their spatial ordering relative to each other and to the carbon skeleton. The localized π electron ordering can be regarded as the formation of two 2D π-electron crystals located on opposite sides of the skeleton (figure 1).

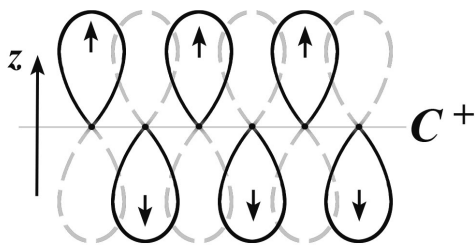

Figure 1. Schematic representation of an atomically flat carbon monolayer consisting of a carbon ion plane and two electron crystals of π electrons in the 2p$_z$ state. For simplicity, 1D crystals are shown instead of 2D ones.

In the case of a hexagonal carbon lattice these electron crystals are characterized by a trigonal symmetry with the distance between π electrons of ~2.46Å and electron density of the order of $10^{15}$cm$^{-2}$. The density is easily calculated for the carbon lattice with a unit cell consisting of two atoms at a distance of ~1.42Å from each other. Note that such a high electron concentration is not consistent with the concept of the energy spectrum linearity which implies that the carrier concentration must be dramatically lower than $10^{16}$ cm$^{-2}$ [19].

Thus, if the Coulomb interaction between electrons is taken into account, the π electron state in the carbon monolayer is characterized by electron crystallization rather than by the 2D electron gas state.

Formation of π bonds is also accompanied by exchange interaction between π electrons. It results in a spin ordering of electrons in both electron crystals: all electron spins are collinear but have opposite directions in two crystals. To put it otherwise, π electron crystals prove to be spin-polarized: spins in each crystal are ordered ferromagnetically, but the ferromagnetically ordered crystals are ordered antiferromagnetically with respect to each other (figure 1). The presence of two electron crystals with different spin polarization directions and their different positions relative to the carbon skeleton plane (on opposite sides) leads to the non-equivalence of neighboring carbon atoms in each monolayer hexagon. Possibly, the STM observation of the trigonal lattice [20] (rather than a hexagonal one) in a single layer of graphite can be regarded as an experimental confirmation of the electron crystal formation.

Since each π electron remains localized at its atom, the energy of its state can change only in a discrete fashion. Therefore, this atomically flat carbon monolayer can be regarded as a quasi-2D mesoscopic quantum system with a sandwich-type structure consisting of a positively charged carbon skeleton plane and two negatively charged planes formed by π electron crystals and parallel to the carbon plane (figure 1).

### 2.2. *Specific features of the quasi-2D mesoscopic quantum system.*

It follows from the π electron localization at atoms and formation of electron crystals that the quasi-2D quantum system has to be a dielectric.

Another feature of this quantum system is associated with specific features of π bonds formed via the resonance of some structures. Though the notion of the resonance state of π bonds (in the sense of the resonance of structures) appeared long ago, the problem of aromaticity of compounds with such bonds and the π electron state in them still remains a matter of controversy [21, 22]. It can be supposed that the formation of three π bonds by one π electron in a flat carbon monolayer occurs via a continuous (or self-sustained) π electron rotation in a circular orbit. The orbit lies in the plane of its π electron crystal and passes through the regions of overlap of its $2p_z$ state with free $2p_z$ states of three neighboring carbon atoms (figure 2) (we recall here that the π electrons of these three neighboring atoms belong to the electron crystal on the other side of the carbon skeleton (figure 1)).

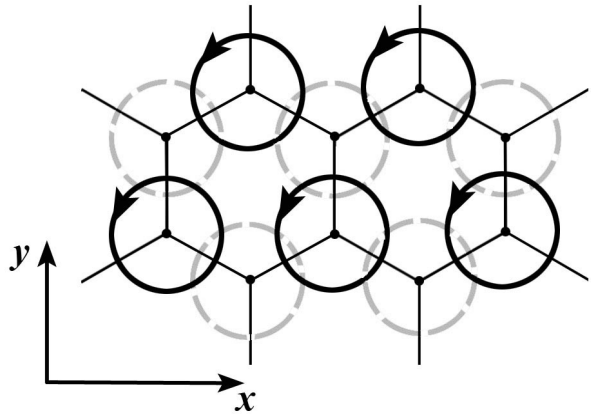

Figure 2. Orbits of self-sustained π-electron rotation in one of the electron crystals (solid lines), the rotation direction (arrows), and free $2p_z$ states of neighboring atoms (dashed lines). The second electron crystal is not shown.

Thus, the π electron remains localized in its atomic state and simultaneously it is rotating in the circular orbit within the bounds of this state. The Coulomb interaction resulting in equal distances between all π electrons in the electron crystal results also in synchronous rotations of electrons in the same direction. The π electron rotations in both crystals should be in synchronism too. The magnetic moments arising in the orbital electron motion can compensate each other if the rotation in the second electron crystal occurs in the opposite direction. Thus, the π electrons in electron crystals of the flat carbon monolayer are ordered not only by the spin, but also by the orbital rotation speed and direction.

Since the electron rotation occurs within the bounds of its atomic state, the rotation speed must be close (by the order of magnitude) to the electron rotation speed in the Bohr atomic model, i.e., it must be thousandths of the light velocity. Perhaps, that is why the experimental cyclotron and band velocities for the carbon monolayer [11, 23] were found to be close (by the order of magnitude) to this well known speed for nonrelativistic particles or, to be more exact, for real electrons. Note that we speak about the electron rotation speed but not about the electron propagation velocity.

Thus, electron crystals in the 2D quantum system that represents an atomically flat carbon monolayer consist of π electrons which synchronously rotate in similar circular orbits lying in the crystal planes. Obviously, such a system must exhibit an appreciable diamagnetism because the electron rotation orbits in both crystals lie in parallel planes.

One more feature of the quantum system we consider is an appreciable electric field gradient on its surfaces. It results from the inability of localized electron crystals to completely screen the positive charge of the carbon skeleton because each electron crystal screens only the charge of the carbon ion to which it belongs and leaves neighboring ions unscreened (figure 3). In other words, the field gradient on the flat carbon monolayer surface is the consequence of the non-equivalence of carbon atoms in the hexagon and a highly asymmetric distribution of electron density in each carbon atom in the $sp^2$ hybridization state. Indeed, the carbon atom has three electrons in the XY plane (in the $sp^2$ state) and only one electron in the Z direction (in one of the $2p_z$ states). Due to the field gradient there is the Coulomb attraction of π electron by three neighboring positively charged carbon ions, which results in its tight binding with them. The binding manifests itself as the π bond formation and leads to the π electron localization, as mentioned above.

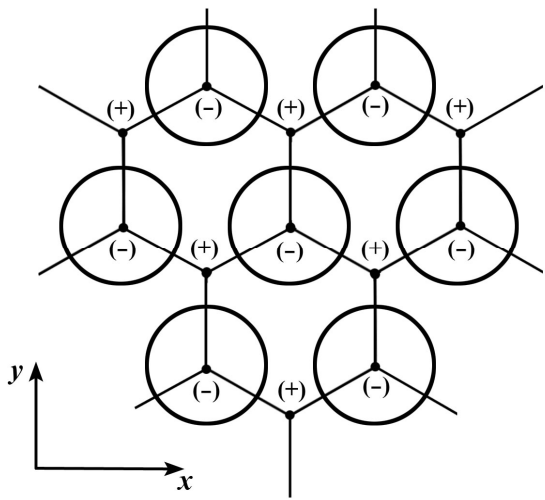

Figure 3. Electric field inhomogeneity and π electron orbits in the electron crystal on one of surfaces of the flat carbon monolayer.

The presence of electron crystals and a highly inhomogeneous electric field on the flat carbon monolayer surfaces leads to an extremely high sensitivity of the quasi-2D mesoscopic quantum system to interactions with external electric fields and charged particles (charges on substrate and contact surfaces, probe tips, etc.) since all the interactions are able to cause distortions in the electron crystals. Moreover, because of a tight binding of π electrons with carbon ions and also their participation in the π bond formation, these distortions are inevitably accompanied by distortions in the carbon skeleton itself.

### 2.3. Distortion of π electron crystals and instability of the 2D quantum system

As an example, we consider the consequences of a point distortion in one of the electron crystals which involves a displacement of one of π electrons from its electron crystal plane due to interaction with an external positive point charge (layer polarization). Such an electron displacement is accompanied by Coulomb repulsion from the remaining π electrons of its electron crystal which pushes the electron out of the crystal in the electron displacement direction. This electron displacement process is irreversible even if the polarization vanishes, since such a polarization, together with the additional repulsive forces, excites the π electron into a higher-energy (relative to its earlier $2p_z$ state) quantum state of the carbon atom and changes its hybridization state.

The quantum state closest to the $2p_z$ state in the carbon atom is a hybrid $2p_z3s$ state. It is difficult to estimate precisely the energy of excitation into this state because the 3s state energy for a carbon atom is unknown. A rough estimate based on the assumption that the 3s state energy of carbon is close to that of a sodium atom calculated in the one-electron approximation [24] yields less than 5 eV. In fact, the excitation energy into the $2p_z3s$ state must be even lower if we take into account the dependence of the state energy on the element atomic number. In any case this estimate shows a high probability of electron excitation into the hybrid state by Coulomb forces.

The carbon ion changes its hybridization state (from $sp^2+2p_z$ to $sp^2+2p_z3s$) and loses the ability to form π bonds with three neighboring ions because its excited hybrid states ($2p_z3s$) can be occupied only by a triplet pair according to Hund's rule. However, the triplet state is in a conflict with the spin polarization of the electron crystals that form π bonds via singlet pairs in the $2p_z$ states (figure 4 and figure 1). Therefore, the bonds between the excited carbon ion and three neighboring ions become single ones (not multiple) and their length increases from 1.42 Å to ~1.50Å. The increase in the bond length forces the excited carbon ion to go out of the carbon skeleton plane, thereby distorting its flatness (figure 4). The change in the bond multiplicity has important consequences for the π electron state in the neighborhood of the excited carbon ion since the electron excitation disrupts the pre-existing balance of Coulomb interactions in both crystals. The establishment of a new balance in the distorted area around the excited ion will be accompanied by changes in multiplicity of bonds between a neighbor and next neighbor (and so on). The neighboring ions also begin to protrude from the carbon plane. In addition, the carbon skeleton curving in the presence of a strong Coulomb interaction between π electrons will also be accompanied by a π electron transition from one crystal to the other (with a subsequent electron excitation into the hybrid $2p_z3s$ state). Evidently, the distorted area expansion, i.e., the transformation of a point distortion into a local one, is completed when the parity of electron densities on both sides of the distorted carbon skeleton is reached, in other words, when the carbon skeleton curvature in the distorted area becomes similar to that of a stable carbon modification with a curved skeleton. Such a modification with a curved carbon skeleton can be a well-known single-wall nanotube ~9 Å in diameter.

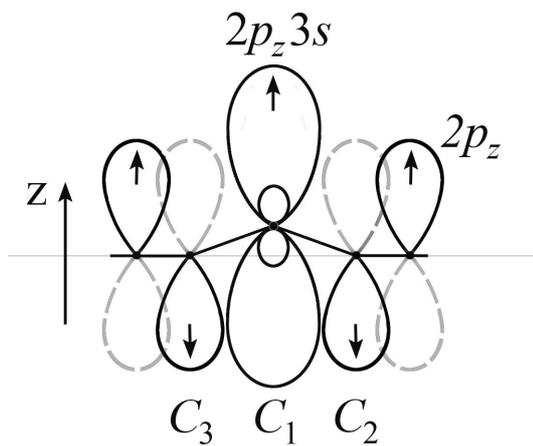

Figure 4. Schematic representation of the first stage of distortion of the carbon monolayer plane by a point charge. The distortion is accompanied by excitation of π electron into the $2p_z3s$ state and formation of single bonds between the excited $C_1$ carbon ion and nonexcited $C_2$ and $C_3$ carbon ions. The third neighboring ion is not shown.

If more than one local distortion occurs in the carbon monolayer, the distortion ordering must take place because it lowers the deformation energy of a charged lattice, both electron and ionic. In the 2D system, ordering via formation of 1D superstructures is the most probable one. As a result, the quasi-2D mesoscopic quantum system with a distorted carbon skeleton can acquire 1D superstructures of excited carbon ions and also additional 1D electron crystals of excited π electrons formed only on the convex side of the quantum system. In the superstructures, ordinary (not related to the resonance of structures) π bonds can be formed by a pair of excited π electrons between two neighboring excited carbon ions. The energy of these π bonds lowers the free energy of the 2D quantum system with the distorted carbon skeleton and contributes to its stabilization. The formation of these bonds implies also that pairing of excited electrons occurs and 1D crystals of excited electrons are substituted by 1D crystals of pairs. Since the excited π electrons have weaker bonds with carbon ions than the π electrons in the $2p_z$ state, it can be supposed that delocalization of the excited electron pairs can take place. However, this will be the delocalization of 1D electron crystals of pairs rather than of individual pairs. The delocalization will occur via tunneling of the 1D electron crystals of pairs through the barriers between the excited states of the π electron pairs and will be strongly anisotropic. The probability of tunneling will depend on the barrier width, i.e., the distance between pairs in 1D electron crystals. It is quite probable that the experimentally observed ballistic conductivity in the carbon nanotubes having a uniformly curved carbon skeleton can be attributed to such a delocalization (nanotubes will be considered elsewhere). It is likely that the experimentally observed low carrier concentration can also be related to these excited electrons or crystals of their pairs.

As to the carbon monolayer, the uniform long-range ordering of numerous local distortions with the curvature similar to that of a single-wall nanotube is highly unlikely (a short-range ordering is more probable). Transformation of the entire monolayer into a set of nanotubes without rupture of σ bonds between carbon atoms is also impossible. Therefore, ordering of numerous local distortions can lead to formation of only fragments of single-wall nanotubes, which makes the carbon monolayer corrugated. This can explain why the corrugated monolayer consists mainly of randomly distributed nanoscale fragments resembling halves of single-wall nanotubes cut along their axis which are arranged alternately with the convex side up and down (thus forming ripples). It is not improbable that the nanoscale fragments which are oriented differently relative to each other can also be interconnected by small fragments similar to fullerene fragments, i.e., the other stable carbon modification with a curved skeleton. Possibly, this explains why the experimentally observed height (~ 10 Å) of intrinsic corrugation is close to the diameter of single-wall nanotube and fullerene [16].

So far we have considered a suspended carbon monolayer. The situation with the substrate-supported monolayer is more complicated because the surface relief is a superposition of the corrugation inherent in the suspended monolayer and that induced by the substrate roughness and Coulomb interaction with the substrate.

It follows from the consideration that properties of a corrugated carbon monolayer are, most likely, determined by the π electron state which is characteristic of stable carbon modifications based on a rolled-up or wrapped-up carbon monolayer rather than an atomically flat one.

## 3. Conclusion

It has been shown that the consideration of the π electron state in the atomically flat 2D carbon monolayer in the approximation of strongly interacting electrons (i.e., taking into account the exchange and strong Coulomb interaction between them) yields the carbon monolayer model differing from that resulting from the idea of weakly interacting particles. In the new approach the atomically flat carbon monolayer is a simplest quasi-2D mesoscopic quantum system consisting of a plane of carbon ions and two planes of electron crystals consisting of spin-polarized orbitally ordered electrons located on opposite sides of the carbon ion plane. The quantum system is characterized by an open electron subsystem and an appreciable electric field gradient on its surfaces.

Localization of π electrons resulting from electron sharing by carbon atoms during π bond formation and also formation of 2D π-electron crystals from rotating (in circular orbits) electrons make this quantum system a dielectric and a diamagnetic with a high magnetic susceptibility.

It has also been shown that this quasi-2D mesoscopic quantum system is extremely unstable and has a tendency to be transformed into more stable wrapped-up or rolled-up modifications. Thus, the carbon monolayer instability is in agreement with the idea of instability of the 2D crystalline lattice [5, 6].

It follows from this consideration that the experimental data obtained for corrugated carbon monolayers characterize properties of a quantum system with a distorted carbon skeleton which contains an additional electron subsystem of excited π electrons on the convex surfaces. The excited π electrons are responsible for chemical and physical properties of such a distorted quantum system. This is in a good agreement with the idea that the unusual features of carbon monolayer can be fully related to the intrinsic ripples in graphene [15].

Though the carbon monolayer as an atomically flat monolayer cannot be long-living its consideration is interesting and useful because it gives an idea of the simplest 2D mesoscopic quantum system with an open electron subsystem. This idea can also contribute to the understanding of the π electron state in fullerenes, carbon nanotubes and graphite. Unfortunately, at present only a qualitative analysis of such a system can be made since calculation of the system is beyond the band-structure description.